# Recent Advances in Diamond Detectors

W. Trischuk and the RD42 collaboration (complete author list at the end)
*Department of Physics, University of Toronto, Toronto, ON, M5S 1A7, Canada*

With the commissioning of the LHC expected in 2009, and the LHC upgrades expected in 2012, ATLAS and CMS are planning for detector upgrades for their innermost layers requiring radiation hard technologies. Chemical Vapor Deposition (CVD) diamond has been used extensively in beam conditions monitors as the innermost detectors in the highest radiation areas of BaBar, Belle and CDF and is now planned for all LHC experiments. This material is now being considered as an alternate sensor for use very close to the interaction region of the super LHC where the most extreme radiation conditions will exist. Recently the RD42 collaboration constructed, irradiated and tested polycrystalline and single-crystal chemical vapor deposition diamond sensors to the highest fluences available. We present beam test results of chemical vapor deposition diamond up to fluences of $1.8 \times 10^{16}$ protons/cm$^2$ showing that both polycrystalline and single-crystal chemical vapor deposition diamonds follow a single damage curve allowing one to extrapolate their performance as a function of dose.

## 1. DIAMOND GROWTH

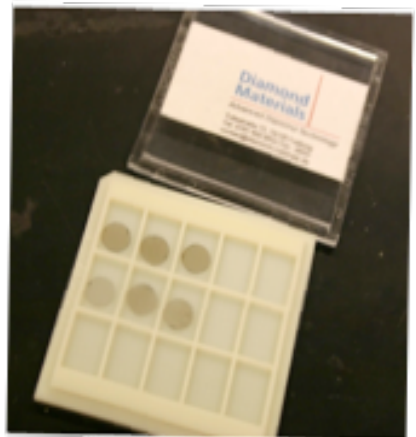

Figure 1: pCVD samples from a new manufacturer. They show collection distances approaching 100 microns.

Its large band-gap and high binding energy make diamond an ideal material for particle detectors in harsh radiation environments. However it is not practical to instrument large areas with natural or high-pressure high-temperature (HPHT) synthetic material. The RD42 collaboration [1], based at CERN, has been working with diamond manufacturers for over a decade to improve the quality of polycrystalline Chemical Vapour Deposited (pCVD) and single crystal Chemical Vapour Deposited (scCVD) diamond material, making it suitable for use as a charged particle sensor. Until recently RD42 has been working with a single manufacturer [2] capable of producing particle detector grade material. Over the last two years a second manufacturer [3] has begun to produce material (see Figure 1) suitable for particle detector prototypes. We are working with this company to ensure that its material has the necessary signal properties, uniformity and radiation tolerance. The appearance of additional manufacturers is encouraging, should lead to cost reductions and additional growth capacity.

## 2. RADIATION TOLERANCE OF DIAMOND SENSORS

Their radiation tolerance is one of the strongest arguments for using diamond sensors in high-energy physics experiments. In addition to retaining significantly more signal than silicon sensors currently deployed in collider experiments, diamond sensors have very low leakage currents – even after large doses of radiation. Since leakage currents dictate the ultimate lifetime of solid-state particle detectors diamond sensors will survive longer at the sLHC.

### 2.1. Testbeam studies

The RD42 collaboration has performed a wide array of radiation tolerance studies of pCVD and scCVD material. This latter material is produced in much the same way as pCVD material, except that it is grown on HPHT diamond substrates. In this way it inherits the ideal crystal structure of the substrate. Unfortunately scCVD samples cannot be grown to sizes much larger than their HPHT templates making them of limited use for particle detection. Still small prototype sensors have been produced, irradiated and tested.

While irradiation studies with reactor neutrons (typically having a few MeV of kinetic energy) and charged pions (with several hundred MeV of momentum) have been carried out, our most mature studies use doses of 24 GeV protons from the CERN PS. Sensors are irradiated in a stand-alone fashion then wire-bonded to strip tracker readout electronics and tested in a charged particle beam. The test setup includes a reference telescope that provides precise (few micron) impact parameter resolution for the reference tracks allowing us to make very precise, background free, determinations of the charge extracted from the irradiated sensors when MIP-like testbeam tracks pass through them.

## 2.2. Universal defect production

We have recently completed studies of 24 GeV proton irradiations that indicate the damage mechanism in pCVD and scCVD material is the same. Figure 2 shows the signal degradation for pCVD (blue points) and scCVD (red points) samples. Shifting the starting point of the fluence scale by -3.8 x $10^{15}$ p/cm$^2$ to their correct un-irradiated collection distance allows us to parametrise the signal degradation by a single fluence ($\phi$) curve: $ccd_{irrad} = \dfrac{ccd_0}{1 + k * \phi * ccd_0}$. Our interpretation is that proton irradiation introduces defects capable of trapping charge in the diamond structure at the same rate in pCVD and scCVD material. The fluence shift required represents the signal advantage of scCVD material. The highest quality pCVD material currently available still includes defects at a density comparable to that introduced into scCVD material by about 4 x $10^{15}$ protons/cm$^2$. From this figure we also see that diamond sensors retain an average signal of 2500 electrons (collection distances in excess of 75 microns) after fluences approaching 2 x $10^{16}$ p/cm$^2$ typical of the doses present at a radius of 4 cm after 10 years running at the sLHC that will have a design luminosity of $10^{35}$ cm$^{-2}$s$^{-1}$.

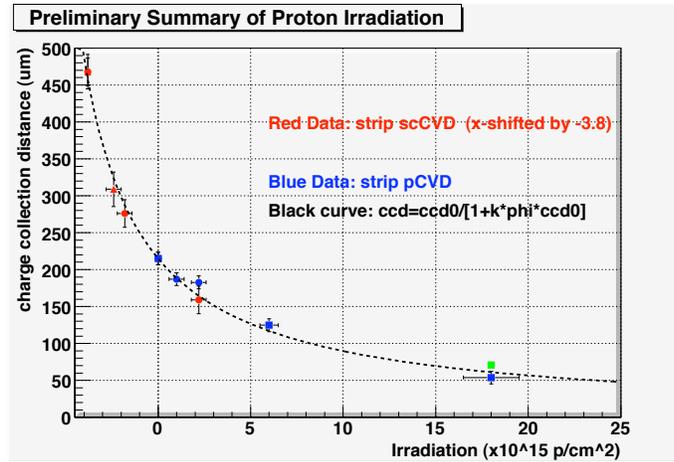

Figure 2: Signals in irradiated pCVD (blue) and scCVD (red) sensors. All points measured at 1V/micron electric field except for the green point taken at 2 V/micron.

## 3. APPLICATIONS OF DIAMOND SENSORS

Diamond sensors have found a widening acceptance in high-energy physics experiments. In this section we describe two current applications and one future application. Following pioneering work by BaBar [4] diamond sensors now form the heart of the CDF beam abort system and have been operational for the past two years. Diamond sensors are also being used in the ATLAS Beam Conditions Monitor [5]. Finally diamond is proposed for the pixel module replacements at the inner radii of the ATLAS sLHC tracker upgrade. Each of these is described briefly in the following.

### 3.1. The CDF beam protection system

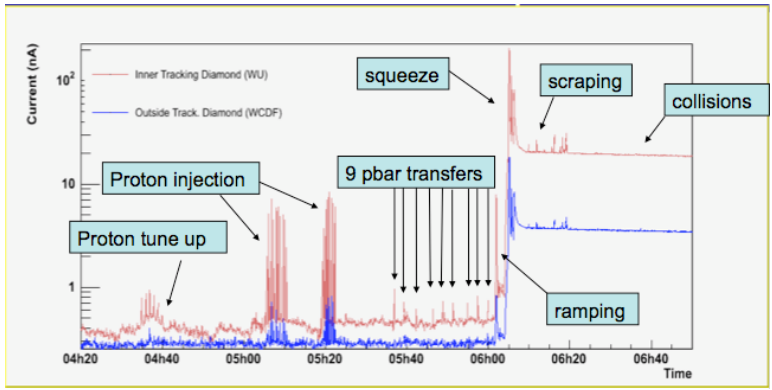

Figure 3: Beam induced current in pCVD diamonds in the inner tracking volume (red) and outside the central tracker (blue) in CDF during beam injection operations at the Tevatron .

An array of 13 pCVD diamond pad sensors – 8 in the inner tracking volume, a distance of less then 2m from the interaction point and 5 additional sensors outside the tracking volume – have been installed in CDF since 2005. They have been monitoring beam operations since early 2006. Figure 3 shows the beam-induced currents seen in these sensors during the injection of beams into the Tevatron. The inner-tracking diamonds (red) are much more sensitive to losses as they are nearer the beam and the final focus for the experiment. Since early 2007 this system has been the default beam protection system for the CDF experiment and has operated reliably correctly anticipating half-a-dozen dangerous beam conditions that were later confirmed by the legacy Tevatron beam loss monitor instrumentation.

### 3.2. The ATLAS Beam Conditions Monitor

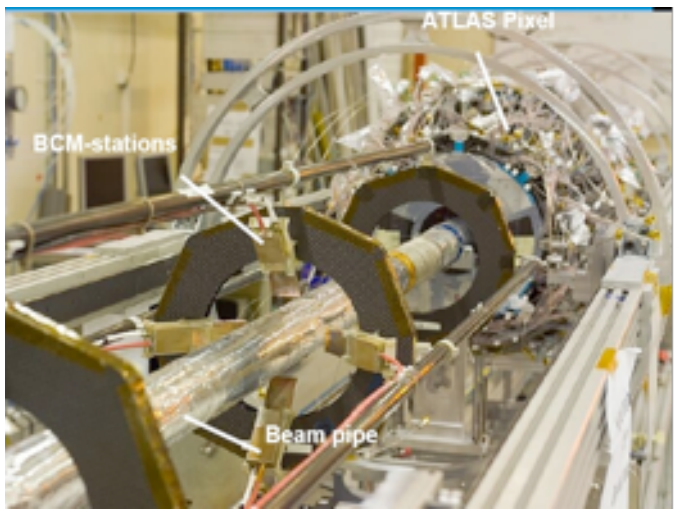

Figure 4: A photo of one end of the ALTAS Beam Conditions Monitor stations. Bunch-by-bunch time-of-flight information is available from each of the four stations.

The ATLAS experiment has recently installed 16 pCVD diamond pad sensors in eight stations – four two metres upstream of the interaction point and four downstream [5]. These sensors, readout in pairs by very fast (2 ns shaping time) radiation tolerant amplifiers, provide time-of-flight information for particles emerging from proton collisions. The time-of-flight for beam-halo particles lost from the LHC beams, is 12.5 ns different from particles produced in collisions, providing optimal distinction from collisions spaced 25 ns apart. The analogue signals are digitised off-detector and arrival time information from the eight modules are compared to those expected for collisions or beam losses by a pair of redundant FPGA processors in the counting-room. This system was installed in the ATLAS experiment in early 2007 and was operational when first beams entered the cavern in the fall of 2008. We are currently refining the beam warning and abort algorithms prior to the first LHC collisions expected in early 2009.

### 3.3. Diamond sensors for the ATLAS sLHC tracker pixels

With the success of diamond sensors in beam monitoring applications and a reliable supply of high-quality sensor material the ATLAS collaboration is considering using diamond in the inner-most layers of its upgraded pixel detector at the sLHC. The R&D necessary to demonstrate the viability of using diamond in this application is outlined in a recently approved proposal [6]. This includes building on the order of 10 pixel modules, thereby taking the first steps towards industrial production of diamond pixel modules; further testing of the position resolution of these modules after exposure to doses comparable to those at the sLHC; and the development of a low-mass support and cooling structure to take advantage of the possibility of running these sensors at room temperature even at the highest doses expected.

## 4. SUMMARY

CVD diamond sensor technology is finding wide spread application in high-energy physics experiments. With the experience gained from systems installed at the Tevatron and the LHC and additional manufacturers diamond is becoming an accepted standard sensor material for high-energy physics. In the years ahead ever more demanding radiation environments, such as those in the inner radii of trackers for the sLHC, will require even more radiation tolerant sensor solutions. The RD42 collaboration continues its study of this material with these applications in mind.

### Authors

The authors of this paper are:

D.Asner[22], M.Barbero[1], V.Bellini[2], V.Belyaev[15], E.Berdermann[8], P.Bergonzo[14], M.Bruzzi[5], V.Cindro[12], G.Claus[10], M.Cristinziani[1], S. Costa[2], R.D'Alessandro[6], W.deBoer[13], I.Dolenc[12], P.Dong[20], W.Dulinski[10], V.Eremin[9], R.Eusebi[7], F.Fizzotti[18], H.Frais-Kolbl[4], A.Furgeri[13], K.K.Gan[16], M.Goffe[10], J.Goldstein[21], A.Golubev[11], A.Gorisek[12], E.Griesmayer[4], E.Grigoriev[11], F.Hugging[1], H.Kagan[16]*, R.Kass[16], G.Kramberger[12], S.Kuleshov[11], S.Lagomarsino[6], A.LaRosa[3], A.LoGiudice[18], I.Mandic[12], C.Manfredotti[18], C.Manfredotti[18], A.Martemyanov[11], M.Mathes[1], D.Menichelli[5], S.Miglio[5], M.Mikuz[12], M.Mishina[7], J.Moss[16], S.Mueller[13], P.Olivero[18], G.Parrini[6], H.Pernegger[3], M.Pomorski[14], R.Potenza[2], S.Roe[3], M.Scaringella[5], D.Schaffner[20], C.Schmidt[8], S.Schnetzer[17], T.Schreiner[4], S.Sciortino[6], S.Smith[16], R.Stone[17], C.Sutera[2], M.Traeger[8], W.Trischuk[19], C.Tuve[2], J.Velthuis[21], E.Vittone[18], R.Wallny[20], P.Weilhammer[3]*, N.Wermes[1].

[1] Universitat Bonn, Bonn, Germany; [2] INFN/University of Catania, Italy; [3] CERN, Geneva, Switzerland; [4] Fachhochschule fur Wirtschaft und Technik, Wiener Neustadt, Austria; [5] INFN/University of Florence, Florence, Italy; [6] Department of Energetics/INFN Florence, Florence, Italy; [7] FNAL, Batavia, U.S.A.; [8] GSI, Darmstadt, Germany; [9] Ioffe Institute, St. Petersburg, Russia; [10] IPHC, Strasbourg, France; [11] ITEP, Moscow, Russia; [12] Jozef Stefan Institute, Ljubljana, Slovenia; [13] Universitat Karlsruhe, Karlsruhe, Germany; [14] CEA-LIST Technologies Avancees, Saclay, Gif-Sur-Yvette, France; [15] MEPHI Institute, Moscow, Russia; [16] The Ohio State University, Columbus, OH, U.S.A.; [17] Rutgers University, Piscataway, NJ, U.S.A.; [18] University of Torino, Italy; [19] University of Toronto, Toronto, ON, Canada; [20] UCLA, Los Angeles, CA, USA; [21] University of Bristol, Bristol, UK; [22] Carleton University, Carleton, Canada. * Spokespersons.